\documentstyle[sprocl]{article}

\def\be{\begin{equation}}
\def\ee{\end{equation}}
\def\bea{\begin{eqnarray}}
\def\eea{\end{eqnarray}}

\def\bbox{{\,\lower0.9pt\vbox{\hrule \hbox{\vrule height 0.2 cm
\hskip 0.2 cm \vrule height 0.2 cm}\hrule}\,}}
\newcommand{\dsl}{\pa \kern-0.5em /}

\def\bbox{{\,\lower0.9pt\vbox{\hrule \hbox{\vrule height 0.2 cm
\hskip 0.2 cm \vrule height 0.2 cm}\hrule}\,}}

\newcommand{\pa}{\partial}

\font\mybb=msbm10 at 12pt
\def\bb#1{\hbox{\mybb#1}}

\def\bR {\bb{R}}

\def\th{\theta}

\def\ti{\tilde }

\input amssym.def
\input amssym.tex
\font\mybb=msbm10 at 10pt
\def\bb#1{\hbox{\mybb#1}}

\def\bR {\bb{R}}

\begin{document}

\begin{flushright}
QMW-PH-99-19\\ hep-th/9911080\\
\end{flushright}
 \vskip 1cm

\title{{DUALITY AND STRINGS, SPACE AND TIME}}

\author{ { C.M. HULL}}

  \address{  Department of Physics\\
  Queen Mary and Westfield College\\
  Mile End Rd, London E1 4NS, UK \\
E-mail: c.m.hull@qmw.ac.uk}

\maketitle\abstracts{
Duality symmetries in M--theory and string theory are reviewed, with particular
emphasis on
the way in which string winding modes and brane wrapping modes can lead to new
spatial dimensions.
Brane world-volumes wrapping around Lorentzian tori can give rise to extra time
dimensions and in this way dualities can change the number of time dimensions
as well as the number of space dimensions.
This suggests that brane wrapping modes and spacetime momenta should be on an
equal footing and M--theory should not be
formulated in a spacetime of definite dimension or signature.}

 \def \R {{\bR}}

\section{String Theory, M-Theory and Duality}

String theory is defined as a perturbation theory in the string coupling
constant $g_s$,
 which  is valid when $g_s$ is small. The fundamental quanta are the
excitations of relativistic strings moving in spacetime
and comprise of   a finite set of  massless particles plus an infinite tower of
massive particles with the scale of the mass
set by the string tension $T=1/l_s^2$,   expressed in terms of a string length
scale $l_s$. If the spacetime  has some
circular dimensions, or more generally has some non-contractible loops, the
spectrum will also include winding modes in which
a closed string winds around a non-contractible loop in spacetime. These have
no analogue in local field theories and are
responsible for some of the key differences between string theories and field
theories. Physical quantities are calculated
through a path integral over string histories, which can be calculated
perturbatively in $g_s$ using stringy Feynman rules,
with string world-sheets of genus $n$ contributing terms proportional to
$g_s^{n}$. In the supersymmetric string theories,
these contributions are believed to be finite at each order in $g_s$,   giving
a perturbatively finite quantum theory of
gravity unified with other forces.

There are {\it five} distinct perturbative supersymmetric finite string
theories, all in $9+1$ dimensions (i.e. nine space
and one time),   the type I, type IIA and type IIB string theories, and the two
heterotic string theories with gauge groups
$SO(32)$ and
$E_8\times E_8$.   The massless degrees of freedom of each of these theories
are governed by a 10-dimensional supergravity
theory, which is the low-energy effective field theory. It has been a
long-standing puzzle as to why there should be five such
theories of quantum gravity rather than one, and this has now been   resolved.
It is now understood that  these are all
equivalent non-perturbatively and that
 these distinct perturbation theories arise as different perturbative limits of
a single underlying theory
\cite{HT,Wit}. We do not have an intrinsic formulation of this underlying
non-perturbative theory yet, but the
relationships between the string theories has been understood through the
discovery of dualities linking them. A central role in
the non-perturbative theory is played by the
$p$-branes. These are $p$-dimensional extended objects, so that a $0$-brane is
a particle, a $1$-brane is a string, a
$2$-brane is a membrane and so on. In the  perturbative superstring theories
there is a 1-brane which is the  fundamental
string providing the perturbative states of the theory, while the other branes
arise as solitons or as D-branes \cite{pol1},
which are branes on which fundamental strings can end. The type II string
theories have a fundamental string and a solitonic
5-brane and a set of D$p$-branes, where $p=0,2,4,6,8$ for the IIA string theory
and $p=1,3,5,7,9$ for the IIB string theory.

There are duality symmetries of string theories that relate brane degrees of
freedom to fundamental quanta, so that all the
branes are on the same footing. If some of the spacetime dimensions are wrapped
into some compact space $K$, so that the
spacetime is $M\times K$ for some $M$, then   branes can wrap around homology
cycles of $K$ and these give extra massive
states in the compactified theory on $M$. For example, a $p$-brane wrapping
around an $n$-cycle with $n \le p$ gives a $p-n$
brane in the compactified theory. These brane wrapping modes generalise the
string winding modes and  are related to the
perturbative states by   U-dualities \cite{HT}, and play an important role in
the duality symmetries, as we shall see.

 One of the best-understood dualities is T-duality \cite{Td}, which relates
string theory on  a spacetime $S^1\times M$ with
a circular dimension of radius $R$
 to a string theory        on  $\tilde S^1\times M$ where the  circular
dimension is now of radius
\be
\tilde R= {l_s^2 \over R}
\ee
 so that the radii $R, \tilde R$ are inversely proportional. For bosonic and
heterotic string theories, T-duality is a
self-duality, so that heterotic (bosonic) string theory on a large circle is
equivalent to heterotic (bosonic) string theory
on a small circle, while it maps the type IIA string theory to  the type IIB
theory, with the result that
 type IIA string theory on a large circle is equivalent to  type IIB string
theory on a small circle \cite{dlp,dsei}.
T-duality relates perturbative states to perturbative states, as does mirror
symmetry which relates a superstring  theory
compactified on a Calabi-Yau manifold $K$ to a superstring  theory compactified
on a topologically distinct Calabi-Yau
manifold, the mirror  $\tilde K$  of $K$.

There are also non-perturbative dualities. For example the type IIA string
theory compactified on $K3$ is equivalent to the
heterotic string theory compactified on the 4-torus $T^4$ \cite{HT}, while the
type I theory with string coupling $g_s$ is
equivalent to the $SO(32)$ heterotic string theory with string coupling $\tilde
g_s = 1/g_s$
\cite{Wit,strstr,dab,witpol}. This is an example of a strong-weak coupling
duality relating the
strong-coupling regime of one theory to the weak-coupling regime of another.
Such dualities are important as they allow the
description of strong-coupling physics in terms of a weakly-coupled dual
theory.

M--theory arises as the strong-coupling limit of the IIA string theory
\cite{Wit}.  The IIA string is interpreted as an
11-dimensional theory  compactified on a circle of radius $R=l_s g_s$.  Then
at strong coupling, the extra   dimension
decompactifies to give a theory in 11 dimensions which has 11 dimensional
supergravity as a low-energy limit. We will refer to
this  10+1  dimensional theory  as M--theory. Duality transformations relate
this to each of the five string theories, and the
string theories and M--theory can all be thought of as arising as different
limits of a  single underlying theory. The IIA
string theory is obtained by compactifying M--theory on a circle,
 the IIB string is obtained from the IIA by T-duality or directly from
compactifying M--theory on a 2-torus and taking the
limit in which it shrinks of  zero size  \cite{Asp}, the $E_8 \times E_8 $
heterotic string is obtained by  modding out
M--theory on a circle by a $Z_2$ symmetry or equivalently from compactifying
M--theory on a line interval \cite{HW}, the type
I theory is obtained from the IIB string by orientifolding (modding out by
world-sheet parity)
\cite{dlp,orientifolds}, and the $SO(32)$ heterotic string is the strong
coupling limit of this \cite{Wit}. The
 type I theory and the $SO(32)$ heterotic string (as well as the type $I'$
string) can be obtained directly from M--theory
compactified on a cylinder as on
 \cite{Bergshoeff:1998re}, while the massive IIA string theory is obtained from
a limit of M--theory compactified on a $T^2$
bundle over a circle  \cite{Hulm}.

In $D$-dimensional general relativity or supergravity, a spacetime with a
large circle $S^1$ is physically distinct from one with a small circle $\tilde
S^1$, and a
spacetime $M\times K$ is physically distinct from the mirror spacetime $M\times
\tilde K$, but in string theory these dual
pairs of spacetimes define the same string theory and so define the same
physics. The heterotic string
on $M\times T^4$ is equivalent to the type IIA string on $M\times K3$,
even though $T^4$ and $K3$ are very different spaces with different properties
(e.g. they have different topologies and
different curvatures) and there is no invariant answer to the question: what is
the spacetime manifold?
In the same way that spacetimes related by
diffeomorphisms are regarded as equivalent, so too must spacetimes related by
dualities, and the concept of spacetime
manifold should be replaced by duality equivalence classes of spacetimes (or,
more generally, duality equivalence classes of
string or M--theory solutions).

In the usual picture, the five superstring theories and the 11-dimensional
theory arising as the strong coupling limit of the IIA
string (referred to as M--theory here) are depicted as being different corners
of the moduli space of  the mysterious
fundamental theory underpinning all of these theories (sometimes also referred
to as M--theory, although we shall resist this
usage here). More precisely,   compactifying string theory or M--theory gives a
theory  depending on the moduli
 of metrics and
antisymmetric tensor gauge fields on the compactification space. Each   modulus
gives rise to a scalar field in the
compactified theory and the expectation value of any of the scalar fields can
be used to define a coupling constant. One can
then examine the perturbation theory in that constant. For some choices it will
give a field theory, for others it will give a
perturbative string theory and different perturbative string theories will
correspond to different choices of coupling
\cite{strong}.  The   string theories and M--theory are each linked to each
other by chains of dualities and so there is only
one basic theory.

More recently,  other \lq corners' corresponding to particular limits of the
theory have been understood to correspond to field
theories without gravity. For example the  IIB string theory in the background
given by the product of 5-dimensional anti-de
Sitter space and a 5-sphere is equivalent to $N=4$ supersymmetric Yang-Mills
theory in four dimensions, with similar results
for theories in other anti-de Sitter backgrounds \cite{mal}, and certain null
compactifications are equivalent to matrix models
\cite{mats}.

Many dualities have now been found which can relate theories with different
gauge groups, different spacetime dimensions,
different spacetime geometries and topologies,
different amounts of supersymmetry, and even relate theories of gravity to
gauge theories. Thus many of the concepts that had
been thought absolute are now understood as relative: they depend on the \lq
frame of reference' used, where the concept of
frame of reference is generalised to include the values of the various coupling
constants. For example, the description of a
given system when a certain coupling is weak can be very different from the
description at strong coupling, and the two
regimes can have different spacetime dimension, for example. However, in all
this, one thing that has remained unchanged is
the number of time dimensions; all the theories considered are formulated in a
Lorentzian signature with one time coordinate,
although the number of spatial dimensions can change. Remarkably, it  turns out
that  dualities can change the number of time
dimensions as well, giving rise to exotic spacetime signatures \cite{sig}. The
resulting picture is that there should be some
underlying fundamental theory and that different  spacetime signatures as well
as different dimensions can arise in various
limits.  The   new theories are different real forms of the complexification of
the original M--theory and type II string
theories, perhaps suggesting an underlying complex nature of spacetime.

We will now proceed to examine some of these dualities in more detail, and in
 particular to focus on the way in which extra spacetime dimensions can emerge
from brane wrapping modes.

\section{Branes and Extra Dimensions}
\subsection{Compactification on $S^1$}

For a {\it field theory} compactified from $D $ dimensions on a circle $S^1_R$
of radius $R$, the  momentum  $p$ in
the circular dimension will  be quantised   with $p=n/R$ for some   integer
$n$. In the limit
$R \to 0$ this becomes divergent, so that      finite-momentum states must move
in the remaining
$D-1$ dimensions and are described by the     dimensionally reduced theory in
$D-1$ dimensions.  For finite $R$, the states
carrying internal momentum  can be interpreted as states in $D-1$ dimensions
with   mass
(taking the $D$-dimensional field theory to be  massless for simplicity)
\be M=\vert n \vert /R
\ee
 The set of all such states for all $n$ gives the  \lq Kaluza-Klein tower' of
massive states  arising from the
compactification. If the original field  theory includes gravity,  there  will
be an infinite   tower of massive gravitons,
and if the theory is supersymmetric, then the tower   fits into supersymmetry
representations. In the limit $R \to 0$, the
masses of all the states in these towers become infinite and they decouple,
leaving the massless dimensionally
 reduced theory in $D-1$ dimensions. On the other hand, taking the
decompactification limit $R \to \infty$, all the states in
the tower become massless and combine with the massless $D-1$ dimensional
fields to form the massless fields in $D $
dimensions. Such a tower becoming massless is often a signal of  the
decompactification of an extra dimension.

For a {\it string theory} the situation is very different, due to the presence
of string {\it winding modes} which become
light as the circle shrinks. A string can wind $m$ times around the circular
dimension, and    the corresponding state in the
$D-1$ dimensional theory will have mass $mRT$ where $T$ is the string tension
(into which a factor of $2 \pi$ has been
absorbed). The set of all such states for all
$m$ forms a tower of massive states and  in the limit $R \to 0$ these become
massless,  so that there is an infinite tower of
states becoming massless (and fitting into supergravity multiplets, in the case
of the superstring). This signals the opening
up of a new  circular dimension of radius
\be
\tilde R= 1/TR
\ee
 with the string winding mode around the original circle of mass
\be M=
 mRT=m/\tilde R
\ee
 reinterpreted as a momentum mode in the dual circle of radius $\tilde R$.
Similarly, the momentum modes on the original
circle  ($M= n/R$) can now be interpreted as string winding modes around the
dual circle ($M=nT \tilde R$), and the new theory
in
$D $ dimensions is again a string theory. A state with momentum $n/R$ and
winding number $m$
will have mass
\be M= {n \over R}+ mRT= {n \over R}+{m \over \ti R}
\ee
and this is clearly  invariant under the T-duality transformation
$ m\leftrightarrow n$, $R\leftrightarrow \tilde R$ interchanging the momentum
and winding numbers and
  inverting the radius. Then the original string theory on $M_{D-1}\times S^1$
(with $M_{D-1}$ some ${D-1}$-dimensional
spacetime) is equivalent to a string theory on $M_{D-1}\times \tilde S^1$ where
$\tilde S^1$ has radius ${\tilde R}$,  with
the momentum modes of one theory corresponding to the winding modes of the
other, and this equivalence is known as a
T-duality. In the limit
$R\to 0,
\tilde R
\to \infty$, the decompactified T-dual theory  has full $D $ dimensional
Lorentz invariance. If the first string theory is a
bosonic (heterotic) string, so is  the second, while if one is a type IIA
string theory, the other is a type IIB string
theory. Then the type IIA string theory compactified on a circle of radius $R$
is equivalent to the type IIB string theory
compactified on a circle of radius $\tilde R$. For 10-dimensional supergravity
compactified on a circle of radius $R$, taking
$R \to 0$ will give a 9-dimensional supergravity theory while for a string
theory a new dimension opens up to replace the one
of radius $R$ that has shrunk to zero size.
If $R$ is much larger than $l_s$, the desciption in terms of string theory on
$S^1 $ is useful,
while for $ R << l_s$, the T-dual description in terms of string theory on
$\tilde S^1 $ is more appropriate, with the
light states having a conventional description in terms of momentum modes on
$\tilde S^1 $ instead of winding modes on $S^1 $.

 \subsection{IIA String Theory and M-Theory}

The type IIA string theory in 9+1 dimensions has D0-brane states, which are
particle-like non-perturbative BPS states, with
quantized charge $n$ (for   integers $n$) and mass
\be M \sim \frac {|n| }{ g_s l_s}
\ee
The state of charge $n$ can be thought of as composed of $n$ elementary
D0-branes. In the strong coupling limit $g_s \to
\infty$, these states all become massless. Moreover, the D0-brane states for a
given $n$ fit into a short massive supergravity
multiplet with spins ranging from zero to two and so at strong coupling there
is an infinite number of gravitons becoming
massless. It was proposed in \cite{Wit} that this tower of massless states
should be interpreted as a Kaluza-Klein tower  for
an extra circular  dimension of radius
\be  R_M= g_s l_s
\ee
 Then the strong coupling limit of the 	IIA string theory is interpreted as the
limit in which $R_M\to \infty$ so that the
extra dimension decompactifies to give a theory in 10+1 dimensions, and this is
  M--theory. Moreover, for the IIA string
theory in $D=10$ Minkowski space, the strong coupling limit is invariant under
the full 11-dimensional Lorentz  group and
the  effective field theory describing the massless degrees of freedom of
M--theory is 11-dimensional supergravity. The radius
can be rewritten in terms of the 11-dimensional Planck length $l_p$ as
\be  R_M= g_s ^{2/3}l_p
\ee

The IIA string theory is really only defined perturbatively for very small
coupling $g_s$. It can now be \lq defined' at
finite coupling $g_s$   as M--theory compactified on a circle of radius $R_M$,
so that the problem is transferred to the one of
defining M--theory. However, at low energies we see that the non-perturbative
IIA theory is described by 11-dimensional
supergravity compactified on a circle, and this   leads to  important
non-perturbative predictions, so that this viewpoint can
be  useful even though we still know rather little about M--theory.

The IIA string has D$p$-branes for all even $p$, while M--theory has a 2-brane
or membrane and a  5-brane. All the branes of
the IIA string theory have an  M--theory origin. For example, an M--theory
membrane will give the fundamental string of the
IIA theory if it wraps around the circular dimension and the D2-brane if it
does not.

\subsection{Compactification on $T^2$}

For a $D$ dimensional field theory compactified on a 2-torus there will be
momentum modes with masses
\be M \sim \sqrt{
\frac {p^2}{R_1^2}+ \frac{ q^2}{R_2^2}}
\ee where $R_1,R_2$ are the radii of the circular dimensions, $p,q$ are
integers and for simplicity we take  the torus to be
rectangular. These will decouple in the limit $R_1,R_2 \to 0$ leaving a theory
in $D-2$ dimensions. For example, for
11-dimensional supergravity, this limit will give the dimensionally reduced
9-dimensional maximal supergravity theory.

We now compare this with   M--theory compactified on $T^2$.  Consider first the
circle of radius $R_2$, say. M--theory
compactified on this circle  is equivalent to
 the IIA string theory with coupling constant $g_s =(R_2/l_p)^{3/2}$, and so
the limit $R_2 \to 0$ is the weak coupling limit
of this IIA string theory. We now have the IIA string theory compactified on a
circle of radius $R_1$, and by T-duality this
is equivalent to the IIB string theory compactified on a circle of radius
$\tilde R_1 = 1/TR_1$. Taking the limit $R_1\to 0$
is then the limit in which
$\ti R_1\to \infty$ and an extra circle opens up  to give the IIB string theory
in 9+1 dimensions. The IIA string winding
modes provide the tower of states that become massless in the limit and which
are re-interpreted as momentum modes on the
circle of radius $\ti R_1$. Moreover, these IIA string winding modes are
M--theory membranes wrapped around the 2-torus. These
membrane wrapping modes have mass
\be M \sim |n| T_2R_1 R_2
\ee  where the membrane tension is $T_2 = 1/l_p^3$.

 Then \cite{Asp} M--theory compactified on a general  2-torus of area $A$ and
modulus $\tau$  is equivalent to the IIB string
theory compactified on a circle of radius
\be R_B= \frac{l_p^3}{A}
\ee with string coupling $g_s$ and axionic coupling    $\th$ (defined as the
expectation value of the scalar field in the
Ramond-Ramond  sector) given by
\be
\tau = \th +i \frac{1}{g_s}
\ee The states of the IIB string carrying momentum in the circular dimension
arise from membranes wrapping the 2-torus
while  the $(p,q)$ string of the  IIB theory winding round the circular
dimension (with fundamental string charge $p$ and
D-string charge
$q$) arises from M--theory states carrying  momentum $p/R_1$ and $q/R_2$ in the
compact dimensions. Then in the limit $A \to
0$, we lose two of the dimensions, as in the field theory, leaving a theory in
8+1 dimensions, but a new spatial dimension
opens up to give a theory in 9+1 dimensions.

\subsection{ Compactification on $T^3$}

For 11-dimensional supergravity compactified on a 3-torus, the limit in which
the radii $R_1,R_2,R_3$ all tend to zero gives
the dimensional reduction to the maximal  supergravity in 8 dimensions. For
M--theory on $T^3$, membranes can wrap any of the
three 2-cycles of $T^3$. From the last section, we know that if
$R_1$ and $ R_2$  both shrink to zero while  $R_3$ stays fixed, an extra
dimension opens up with radius $\ti R_3 =
l_p^3/R_1R_2$ and the tower of membrane wrapping states is reinterpreted as a
Kaluza-Klein tower for this extra dimension. The
same picture   applies to each of the three 2-cycles, and so  if all 3 radii
shrink, there are three extra dimensions  opening
up, with radii $\ti R_i$ given by
\be
 \ti R_i = \frac {l_p^3}{R_jR_k}, \qquad  i\ne j \ne k
\label{elv}
\ee Then when the original three torus shrinks to zero size, three dimensions
are lost but three new ones emerge, so we are
again back in 11 dimensions and the  11-dimensional theory is again M--theory
\cite{fhrs}. Thus M--theory compactified on a
dual  $T^3$ with radii $R_1,R_2,R_3$ is equivalent to M--theory compactified on
the dual  $T^3$ with radii $\ti R_1,\ti R_2,\ti
R_3$ given by (\ref{elv}).

\subsection{ Compactification on $T^4$}

It is tempting to apply these arguments  to higher tori. For example, $T^4$ has
six 2-cycles, and membranes can wrap any of
them. Compactifying $D=11$ supergravity  on a $T^4$  and taking all four radii
$R_i \to 0$ gives a $D=7$ field theory, but for
M--theory there are six towers of states becoming massless in the limit arising
from membranes wrapping each of the six
shrinking 2-cycles. If each of these towers is interpreted as a Kaluza-Klein
tower, this would give
  6 extra dimensions in addition to the 7  original dimensions remaining,
giving a total of 13 dimensions. However, there is
no conventional   supersymmetric theory in 13 dimensions, so it is difficult to
see how such a theory could emerge. In fact
the situation here is more complicated, and the 6 towers have a different
interpretation here. The difference here is that
there is also a string in the compactified theory  arising from the M--theory
5-brane wrapped around the $T^4$ which becomes
\lq light' at the same time as the 6 towers of membrane wrapping  modes
\cite{strong}. It turns out that M--theory on $T^4$
is dual to IIB string theory on $T^3$. The M--theory 5-brane wrapped around the
$T^4$ gives the fundamental string of the  IIB
theory moving in 7 dimensions \cite{strong}. Compactifying the IIB string on
$T^3$ gives in addition three momentum modes and
three winding modes, fitting into a {\bf 6} of the T-duality group
$SO(3,3)$, and these correspond to the 6 towers of membrane wrapping modes,
which themselves transform as a   {\bf 6} of the
torus group
$SL(4) \sim SO(3,3)$. Thus only 3 of the 6 towers can be interpreted as
momentum modes for an extra dimension, the other three
being interpreted as string winding modes, and the spacetime dimension of the
dual theory is 10, not 13. Note that there is no
invariant way of choosing which three of the six correspond to spacetime
dimensions, as T-duality transformations will relate
momentum and winding modes and change one subset of three to another. In this
case, taking the limit in which the $T^4$ on
which the M--theory is compactified  shrinks to zero size does not correspond
to a decompactification limit   of the dual
theory, but to the weak coupling  limit in which the coupling constant $g_s$ of
the  compactified IIB string theory tends to
zero \cite{strong}.

\section{Branes and Space and Time}

We have seen that wrapped branes are associated with towers of  massive states
and that in some cases these can be interpreted
as Kaluza-Klein towers for extra dimensions. In a limit in which such a   tower
becomes massless (e.g. $R_i\to 0$ for toroidal
compactifications, or $g_s \to \infty$ for the IIA D0-branes), the
corresponding dimension decompactifies and new dimensions
unfold. The presence of an enlarged Lorentz symmetry puts  the new braney
dimensions on an equal footing with the other
dimensions, and the full theory includes gravity in the enlarged space. The
number of dimensions lost in the limit is not
always the same as the number of extra dimensions, so that the total number of
spacetime dimensions can change (as in the
relations between 11-dimensional M--theory and 10-dimensional type II string
theories considered in sections 2.2 and 2.3).
 We have also seen that, as in the case of M--theory on $T^4$,  the towers of
wrapped brane states cannot always be interpreted
in terms of extra dimensions, and it is necessary to perform a more complete
analysis to see what is going on.

In all of the above cases, branes were wrapped around spacelike cycles and the
extra dimensions that arose were all spacelike.
A brane world-volume can also wrap around timelike cycles, and we will see that
in such cases the extra  dimensions can be
timelike, so that the signature of spacetime can change.

It is natural to ask whether it makes sense to consider compact time. There are
many classical solutions of gravity,
supergravity, string and M--theories with compact time and it is of interest to
investigate their properties. Compact time does
not appear to be a feature of our universe, but almost all spacetimes that are
studied  are also unrealistic.The presence of
closed timelike loops means that the physics in such spaces is unusual, but
 it has often been fruitful in the past to  study solutions that have little in
common with the real world. An important issue
with these solutions (as with many others) is whether a consistent quantum
theory can be formulated in such backgrounds.  If
time were compact but with a huge period, it is not clear how that would
manifest itself.

With a compact time, it is straightforward to solve classical field equations,
imposing  periodic boundary conditions in time
instead of developing Cauchy data. Can quantum theory make sense with compact
time? There is no problem in solving
Schr\" odinger or wave equations with periodic boundary conditions, but it is
difficult to formulate any concept of measurement
or collapse of a wave-function, as these would be inconsistent with periodic
time: if a superposition of states collapsed to an
eigenstate of an observable in some measurement process, it must already have
been in that eigenstate from the last time it
was measured. In string theory, it is straightforward to study the solutions of
the physical state conditions, but there are
new issues that arise from string world-sheets (and brane world-volumes)
wrapping around the compact time. It has proved very
fruitful to consider such compactifications in string theory. For example, the
compactification of all 25+1 dimensions in
bosonic string theory on a special Lorentzian torus played a central role in
the work of Borcherds on the construction of
vertex algebras and their application to the monster group \cite{bor}.

\section{Compactification on Lorentzian Tori and Signature Change}

\subsection{Compactification on a Timelike Circle}

Consider spacetimes of the form $M_{D-1}\times S^1 $ where $S^1$ is a timelike
circle of radius $R$ and $M_{D-1}$ is a
Riemannian space. The time component of momentum is quantized
\be p^0=\frac n R
\ee and in the limit $R\to 0$, only the states with $p^0=0$ survive. For a
field theory, the result is  a dimensional
reduction to a Euclidean field theory in $D-1$  dimensions, on $M_{D-1}$.  For
example, dimensionally reducing $D=11$ supergravity on a
timelike circle gives a supergravity theory in 10 Euclidean dimensions, denoted
the $IIA_E$ supergravity theory in \cite{sig}.
Timelike reductions of supergravity theories have been considered in
\cite{HJ,CPS,Stelle}.
The field theory resulting from such a timelike reduction will in general have
fields whose kinetic terms have the wrong sign. For example,
the $D$ dimensional graviton will give  a graviton, a scalar and a
  vector field in $D-1$ dimensions  on reducing on a circle, and if the circle
is timelike,
then the vector field will have a kinetic term of the wrong sign.
Then the action for the physical  matter fields of the reduced theory in $D-1$
Euclidean dimensions will not be positive.
This apparent problem is the result of the truncation to $p^0=0$ states.
If this truncation is not made and if the
full Kaluza-Klein towers of states with $p^0=  n/ R$ for all $n$ are kept, then
the theory is the full unitary $D$ dimensional
theory on a particular background, and
 the $D$ dimensional gauge invariance can be used to choose a physical gauge
locally with a positive action and  states with positive norm.    In general
such a gauge choice cannot be made globally and there will be zero-mode states
for which the action will not be positive.
For example, the states with $p^0=0$ are governed by the non-positive
dimensionally reduced action in $D-1$ dimensions.
For a Yang-Mills theory reduced  on a timelike circle,  the
time component of the vector potential $A_0$ gives a scalar field in $D-1$
dimensions with a kinetic term of the wrong sign.
For the full $D$ dimensional theory compactified on the timelike circle, the
negative-norm
$A_0$
 can be brought  to a constant  by $D$-dimensional  gauge transformations, but
one cannot gauge away the
degrees of freedom associated with Wilson lines winding around the compact time
dimension.
(The fields with kinetic terms  of the wrong sign can be handled in the path
integral in the same way as the negative-action
gravitational conformal mode is sometimes dealt with, namely by analytic
continuation so that the offending field becomes
imaginary \cite{ebr}.)

In a string theory, however, there will be winding modes in which the 1+1
dimensional string world-sheet  winds around the
compact time dimension, giving a spacelike \lq world-line' in the compactified
theory in $D-1$ dimensions. As in the spacelike
case, as $R\to 0$, a  dual circle opens up with radius $\ti R=1/TR$, and the
new circle is again timelike. The winding number
becomes the $p^0$ of the dual theory, and in this way  a superstring theory in
9+1 dimensions compactified on a timelike
circle of radius $R$ is T-dual to a superstring theory in 9+1 dimensions
compactified on a timelike circle of radius $\ti R$.
Such timelike T-dualities were considered for the bosonic and heterotic strings
in e.g. \cite{Moore}, and they take the bosonic
string theory to the bosonic string theory  and the heterotic string theory  to
the heterotic string theory. However, for type II theories there is a
surprise. It is straightforward to see that timelike T-duality cannot take the
IIA string theory to either the IIB string or
the IIA string \cite{ebr}, but must take it to a \lq new' theory, denoted the
$IIB^*$ string theory in \cite{ebr}. Similarly,
timelike T-duality takes the IIB string to a   $IIA^*$ string theory
\cite{ebr}.

The $IIA^*$ and $IIB^*$ strings are taken
into each other by  T-duality on a spacelike circle, and the $IIA^*$ ($IIB^*$)
theory is obtained from the IIA (IIB) string
theory by acting with
$(i)^{F_L}$  \cite{ebr} (where $F_L$ is the left-handed fermion number).
The supergravity limits of the  $IIA^*$ and $IIB^*$
have non-positive actions for the matter fields (the kinetic terms for the
fields in the R-NS and R-R sectors have the wrong
sign)
so that the low-energy field theories are non-unitary, but  the  $IIA^*$ and
$IIB^*$ string theories compactified on a
timelike circle are  equivalent to the IIA and IIB string theories on the dual
timelike circle.
Then, at least when on a  timelike circle, the  $IIA^*$ and $IIB^*$ string
theories
are precisely the timelike compactifications of the usual IIA and IIB string
theories, albeit
 written in dual variables.  The supergravity limit
for the IIA or IIB variables is the conventional one, while the supergravity
limit for  the dual variables is non-unitary.  A
physical gauge can then be chosen locally for the
$IIA^*$ and $IIB^*$ string theories
on a timelike circle, and any lack of unitarity or positivity is due to
zero-modes.

If time is compact and the physics is periodic in time, the requirements for a
sensible theory are not the same as in Minkowski
space. A theory that is unstable in Minkowski space (perhaps due to negative
energy configurations)
need not be pathological if time is compact: the periodic boundary conditions
forbid any
runaway solutions and the system will always return to its starting point after
a period.
A nonunitary theory in Minkowski space will not conserve probability, but with
periodic time, any probability that is lost will always come back, as the
solutions of the wave equations are required to be periodic. This suggests that
the timelike compactifications of the  $IIA^*$ and $IIB^*$ string theories
should be consistent, although the question remains as to the status of the
decompactification limit in which the radius of the timelike circle becomes
infinite. Similar considerations will apply to the other new theories of
\cite{sig} described in this section.
 See \cite{ebr} for further discussion of the type $II^*$ theories.

\subsection{Compactification on $T^{1,1}$}

Consider now compactification on the Lorentzian torus  $T^{1,1}$ with one
spacelike circle    and one timelike one. (We will
use the notation $T^{s,t}$ for a torus with $s$ spacelike circles    and $t$
timelike ones.) For 11-dimensional supergravity,
the limit $R_s,R_t \to 0$ gives a 9-dimensional Euclidean supergravity theory.
For M--theory on a Euclidean torus $T^2$, we saw
in section 2.3 that in the limit in which the torus shrank to zero size, one
new spacelike dimension opened up to give the IIB
string theory in 9+1 dimensions.  Here we expect something similar to happen.
Considering first the compactification  on  the
spacelike circle  of radius $R_s$,  when $R_s$ is small we obtain the IIA
string theory with coupling constant
$g_s=(R/l_p)^{3/2}$. The compactification of this on a timelike circle of
radius $R_t$ is T-dual to the $IIB^*$ string theory
compactified on a timelike circle of radius
\be
\tilde R_t = {1	 \over TR_t}
\ee Then taking the limit $  R_t \to 0$, we obtain a theory in the expected 9
spacelike dimensions together with a new time
dimension which opens up, the T-dual of the original timelike dimension. The
membranes wrapping around $T^{1,1}$ have become
the  modes carrying the time component of momentum $p^0$ of the dual $IIB^*$
theory, and   M--theory compactified on $T^{1,1}$
 with radii $R_s,R_t$ is dual to the $IIB^*$ string theory compactified on a
timelike circle of radius $l_p^3/R_sR_t$, as was
shown in  \cite{sig}.

\subsection{Compactification on $T^{2,1}$}

We have seen in section 2.3 that M--theory compactified on a Euclidean 2-torus
$T^2$  gains a new spatial dimension in the limit
in which the 2-torus shrinks to zero size, replacing the two which have
disappeared, so that the original theory in (10,1)
dimensions  becomes a theory in (9,1)  dimensions: $(9,1)=(10,1)-(2,0)+(1,0)$.
Similarly, we have seen in section 4.1 that
M--theory compactified on a Lorentzian 2-torus $T^{1,1}$  gains a new time
dimension in the limit in which the 2-torus shrinks
to zero size, replacing the $(1,1)$ dimensions  which have disappeared so that
the original theory in (10,1) dimensions  again
becomes a theory in (9,1) dimensions: $(9,1)=(10,1)-(1,1)+(0,1)$. Thus a
shrinking $T^2$ is associated with an extra space
dimension while a shrinking $T^{1,1}$ is associated with an extra time
dimension.

For M--theory on a shrinking Euclidean $T^3$, an extra space dimension emerges
for each of the three shrinking 2-cycles, so that
the three toroidal dimensions which are lost are replaced by three new spatial
dimensions,  and we end up back in M--theory in
(10,1) dimensions: $(10,1)=(10,1) - 3 \times (1,0)+3 \times (1,0)$.

Consider now the compactification on a Lorentzian 3-torus $T^{2,1}$ with two
spacelike and one timelike circles. In the limit
in which the torus shrinks to zero size, 2+1 dimensions are lost leaving 8
Euclidean dimensions and reducing 11-dimensional
supergravity on $T^{2,1}$ indeed gives a supergravity in (8,0) dimensions. In
M--theory, if the discussion above applies here,
we expect an extra space dimension for every shrinking $T^2$ and an extra time
dimension for every shrinking $T^{1,1}$. The
torus $T^{2,1}$ has two Lorentzian 2-cycles and one Euclidean one, so that this
suggests there should be an extra two time
dimensions and one space dimension that open up in this limit, giving a theory
in 11 dimensions with two-timing signature
$(9,2) = (8,0)+(1,0) +2 \times (0,1)$. If all the towers of wrapped membranes
give extra dimensions, this must be the result,
but we have seen that in some cases towers of wrapped brane states can have
other meanings. A more careful analysis shows that
  this interpretation is indeed correct and taking M--theory on a shrinking
$T^{2,1}$   gives a new theory in 9+2
dimensions \cite{sig}.

Then dualities can change the number of time dimensions as well as the number
of space dimensions. This new theory in 9+2
dimensions was referred to as  the $M^*$ theory in \cite{sig}, and  it has an
effective field theory which is a new
supergravity theory in 9+2 dimensions. M--theory compactified on $T^{2,1}$ is
equivalent to $M^*$ theory compactified on  a
two-time torus $T^{1,2}$, with the sizes of the circles related by a formula
similar to (\ref{elv}).

\subsection{Compactifications of $M^*$ Theory}

We can now investigate the compactifications of $M^*$ theory on various tori
\cite{sig}.  Compactifying the $M^*$ theory on a
timelike circle gives the $IIA^*$ string theory in 9+1 dimensions, while
compactifying on a spacelike circle gives a new
IIA-like string theory in 8+2 dimensions. Next consider the compactification on
2-tori  in the limit in which they shrink to
zero size. For $T^{0,2} $ this gives the $IIB$ string (compactification on the
first circle gives the $IIA^*$ theory and the
second then gives its T-dual on a timelike circle), for $T^{1,1}$ it gives the
$IIB^*$ theory and for $T^{2,0}$ it gives a new
IIB-like theory in 7+3 dimensions. Thus a shrinking $T^{0,2} $ gives an extra
time dimension,  a shrinking $T^{1,1} $ gives
an extra space dimension and  a shrinking $T^{2,0} $ gives an extra time
dimension. This can now be used to find the results of
compactification on a shrinking three-torus. For $T^{1,2}$ there are two
$T^{1,1}$ cycles and one Euclidean  $T^2$ cycle
giving a theory in $(9,2)-(1,2)+ 2\times (1,0) +(0,1)=(10,1)$ dimensions and we
are back in M--theory, for
 $T^{2,1}$ there are two $T^{2}$ cycles and one    $T^{1,1}$ cycle giving a
theory in $(9,2)-(1,2)+ 2\times (0,1)
+(1,0)=(9,2)$ dimensions and we are back in $M^*$ theory, while  for
$T^{3,0}$ there are three Euclidean  $T^2$ cycles giving a theory in
$(9,2)-(3,0)+ 3\times (0,1)=(6,5)$ dimensions, giving a
new theory in 6+5 dimensions. This theory was denoted the $M'$ theory in
\cite{sig}, and $M^*$ theory compactified on
$T^{3,0}$ is equivalent to the $M'$ theory compactified  on  a dual $T^{0,3}$.

The above analysis can then be repeated for this new  $M'$ theory, and it turns
out that  only 11-dimensional theories that
arise are the
  $M,M^* $ and   $M'$ theories, with  signatures  (10,1), (9,2) and (6,5),
together with
the  mirror theories in signatures (1,10), (2,9) and (5,6). Reduction on
circles gives IIA-like theories in signatures 10+0,
9+1, 8+2, 6+4 and 5+5 while reducing on 2-tori gives IIB-like theories in
signatures  9+1, 7+3,
  and 5+5. (There are of course also mirror string  theories in the signatures
1+9, 2+8 etc with space and time interchanged.)

In each of these 10 and 11 dimensional cases there is a corresponding
supergravity limit  and it is a non-trivial result that
these supergravities exist, and it is unlikely that there are maximal
supergravities in signatures outside this list.  These
theories are linked to each other by an intricate web of dualities \cite{sig},
some of which have been outlined above, and in
particular all are linked by dualities to M--theory.

Each of these theories has a set of branes of various world-volume signatures
\cite{khur,khura}. For the M--type
theories, M--theory has branes of world-volume signature 2+1 and 5+1 (the usual
M2 and M5 branes), $M^*$ theory  has branes of
world-volume signature 3+0,1+2 and 5+1 while
$M'$ theory  has branes of world-volume signature 2+1,0+3, 5+1, 3+3 and 1+5.

\section{Discussion}

In a field theory, compactification and then shrinking the internal space $K$
to zero size gives a dimensionally reduced field  theory in lower dimensions.
In compactified  string theory or M--theory, however,
new dimensions can emerge when the internal space shrinks, with the
Kaluza-Klein towers
for the new dimensions corresponding to the brane wrapping modes in which
branes
wrap around cycles of  $K$. In some cases (e.g. toroidal compactifications of
string theory
or M--theory on $T^3$) the number of new dimensions equals the number that are
lost and one regains the original spacetime dimension, while in others (such as
M--theory compactified on $T^2$) the number of new dimensions is different from
the number that are lost and so the dimension of spacetime changes (for
M--theory on $T^2$ it changes from 11 to 10).

Clearly, the notion of what is a spacetime dimension is not an invariant
concept, but depends on the \lq frame of reference', in the sense that it will
depend on the values of various moduli.
 A given tower of BPS states could have a natural interpretation as a
Kaluza-Klein tower
associated with momentum in a particular compact spacetime dimension for one
set of parameters, but could have  an interpretation as a tower of brane
wrapping modes for other values, and we have seen many examples of this in the
preceding sections.
We are used to considering field theories in  spacetimes of given dimension and
signature,
but any attempt to formulate M--theory or string theory as a theory in a given
spacetime dimension or signature will be misleading.
In particular,  the theory underpinning
all these theories has a limit which behaves like a theory in 10+1 dimensions
with a supergravity limit and systematic corrections, but cannot at the
fundamental level be a theory in 10+1 dimensions, as it has some limits which
live 9+1 dimensions and others that live in 9+2 or 6+5 dimensions.

The   supersymmetry algebra in 10+1 dimensions
  is
\be
\{Q,Q\} = C(\Gamma^{M}P_{M} - {1\over 2!}\Gamma^{M_1 M_2}Z_{M_1 M_2} -
{1\over 5!}\Gamma^{M_1\dots M_5}Z_{M_1\dots M_5})\ ,
\ee
where $C$ is the charge conjugation matrix, $P_{M}$ is the
energy-momentum 11-vector and $Z_{M_1 M_2}$ and $Z_{M_1\dots M_5}$ are
2-form and 5-form charges, associated with brane charges \cite{Hull}.  There
are $11$+$55$+$462$=$528$
 charges on the right-hand-side, which can be assembled into a symmetric
bi-spinor $X_{\alpha \beta}$. Compactifying and then dualising, one finds that
some of the brane charges become momenta of the dual theory and some of the
momenta become brane charges of the dual theory, so that the split of the
bi-spinor $X$  into an 11-momentum
and brane charges changes under duality.

This suggests that
rather than trying to formulate the theory in 10+1 dimensions, all 528 charges
should be treated in the same way. There seem
to be at least two ways in which this might be done. The first would be a
geometrical one in which all 528 charges were
treated as momenta and there is an underlying spacetime of perhaps 528
dimensions. The duality symmetries could then act
geometrically, and there would be perhaps some dynamical way of choosing 11 of
the dimensions as the preferred ones, e.g.
through the \lq world' being an 11-dimensional surface in this space. For
example, in considering T-duality between a string
theory on a space $M\times S^1$ and one in the dual space $M\times \tilde S^1$,
it is sometimes
 useful to consider models on $M\times S^1\times \tilde S^1$ in which both the
circle of radius $R$ and the dual circle of
radius $ \tilde R$ are present, with different projections or gaugings giving
the two T-dual models; see \cite{Td} and
references therein.

We have seen that different spacetimes related by dualities can define the same
physics, so that the notion of spacetime
geometry cannot be fundamental.
This suggests that different degrees of freedom should be used, with spacetime
emerging as a derived  concept.
An alternative \lq anti-geometrical' formulation would be one in which none of
the charges were
geometrical, but instead an algebraic approach similar to that of matrix theory
was used. For example, M--theory could be
compactified to 0,1 or 2  dimensions to give a theory that would be expected to
have  duality symmetry \cite{Julia}
 $E_{11},E_{10}$ or $E_{9}$
where $E_9$ is an affine $E_8$, $E_{10}$ is a hyperbolic algebra discussed, for
example, in \cite{eten} and $E_{11}$ might be  some huge algebraic structure
associated with the
 $E_{11}$ Dynkin diagram. In one dimension the theory might be some matrix
quantum mechanics associated with
$E_{10}$ while in zero dimensions it would be some form of  non-dynamical
matrix theory.  At special points in the moduli
space, some of the  charges would be associated with extra dimensions that are
decompactifying. At different points, different
numbers of space and time dimensions could emerge.

Such formulations
might be related  to the reformulations of 11-dimensional supergravity of
\cite{nic,nic1,nic2}
in which the tangent space group is enlarged so that
some of the duality symmetries are manifest. For example, in the context of
compactifications to 2+1 dimensions,
the usual tangent space group $SO(10,1)$ is    broken to $SO(2,1) \times SO(8)$
and then anti-symmetric tensor degrees of
freedom were used in \cite{nic1}  to reformulate the theory with tangent space
group
 $SO(2,1) \times SO(16)$, with the $SO(16)$ associated with the usual local
$SO(16)$ invariance of  3-dimensional supergravity. These formulations show
that there are
alternatives to the usual formulation in 11 spacetime dimensions and it would
be interesting to consider others.

The five superstring theories and   M--theory   are  different corners of the
moduli space of  some as yet unknown
fundamental theory and the   dualities linking them all involve
compactification on Riemannian spaces. If
this is extended to include compactification on spaces with Lorentzian
signature
 a richer structure emerges.  The strong coupling limit of the type IIA
superstring is M--theory in 10+1 dimensions whose low
energy limit is 11-dimensional supergravity theory. The type I, type II and
heterotic superstring theories and  certain
supersymmetric gauge theories   emerge as different limits of M--theory. The
M--theory in 10+1 dimensions is  linked via
dualities to $M^*$ theory in 9+2 dimensions  and $M'$-theory in 6+5 dimensions.
Various limits of these   give rise to
IIA-like string theories in 10+0, 9+1,8+2,6+4 and 5+5 dimensions, and to
IIB-like string theories in   9+1,7+3,  and 5+5
dimensions. The field theory limits  are supergravity theories with 32
supersymmetries in  10 and 11 dimensions with these
signatures, many of which are new. Further dualities similar to those of
\cite{mal} relate these
  to supersymmetric gauge theories in various signatures and dimensions, such
as 2+2, 3+1 and 4+0.
These  new string theories and M--type theories in various spacetime signatures
can all be thought
of as providing  extra corners   of the moduli space.
Some corners are stranger than others, but in any case we can only live in one
corner (perhaps M--theory compactified on the
product of a line interval  and a Calabi-Yau 3-fold) and there is no reason why
other corners might not have quite unfamiliar
properties.

Theories in non-Lorentzian  signatures usually have many problems, such as lack
of unitarity and instability.
However,   the theories considered here  are related to M--theory via dualities
and so are just the usual theory expressed in
terms of unusual variables. For example, the $M^*$ theory in 9+2 dimensions
compactified on $T^{1,2}$
is equivalent to M--theory compactified on $T^{2,1}$, and so the compactified
$M^*$ theory will make sense provided
M--theory compactified on a Lorentzian torus is a consistent theory. Then the
problems with formulating a theory in
9+2 dimensions are   in this case only apparent, as the theory can be rewritten
as  a theory in 10+1 dimensions using
different variables, so that the extra time dimension is replaced by the
degrees of freedom associated with   branes wrapped
around time.

There are several possible generalisations of the notion of a particle to
general signatures.
A physical  particle or an observer in Lorentzian spacetime with signature
$(S,1)$ follows a timelike (or null) world-line
while a tachyon would follow a spacelike one.
In a space of signature $(S,T)$, one can again consider worldlines of signature
$(0,1)$, but  other generalisations of particle
might include branes with worldvolumes (\lq time-sheets')
 of signature $(0,t)$ with $t \le T$, sweeping out some or all of the times.
In a general signature $(S,T)$, it is natural to consider branes of arbitrary
signature $(s,t)$ with $s\le S$ and $t \le T$,
and the conditions on $(s,t)$ for these to be supersymmetric were given in
\cite{khur}.

In conclusion, we have reviewed part of the intricate web of duality symmetries
linking many apparently different theories, but
since the theories are all related in this way, they should all be regarded as
corners of a single underlying theory.
In particular, two dual theories can be formulated in spacetimes of different
geometry, topology and even signature and
dimension, and so all of these concepts must be relative rather than absolute,
depending on the values of certain parameters
or couplings, and such a relativity principle should be a feature of
  the fundamental theory that underlies all this.



\section*{References}
\begin{thebibliography}{99}


\bibitem{HT} C.M.~Hull and P.K.~Townsend,
Nucl.\ Phys.\ {\bf B438} (1995) 109 hep-th/9410167.


\bibitem{Wit} E.~Witten,
{Nucl.~Phys.}~{\bf B443} (1995) 85, {hep-th/9503124}


\bibitem{pol1}
J.~Polchinski,
 {hep-th/9611050}


\bibitem{Td} A. Giveon, M. Porrati and E. Rabinovici, Phys. Rep. {\bf 244}
(1994) 77.

\bibitem{dlp} J.~Dai, R.G.~Leigh and J.~Polchinski,
Mod.~Phys.~Lett. {\bf A4}
(1989) 2073;

\bibitem{dsei} { M. Dine, P. Huet and N. Seiberg, Nucl. Phys. {\bf B322} (1989)
301.}


\bibitem{strstr} C.M.~Hull,
Phys.\ Lett.\ {\bf B357} (1995) 545, hep-th/9506194.


\bibitem{dab}{A. Dabholkar, Phys. Lett. {\bf B357} (1995) 307, hep-th/9506160.}

\bibitem{witpol}{J. Polchinski and E. Witten,
Nucl.Phys. {\bf B460} (1996) 525,
hep-th/9510169.}


\bibitem{Asp} P.~Aspinwall,
Nucl. Phys. Proc. Suppl. {\bf  46}
(1996) 30,
 {hep-th/9508154} ;\\
J.H.~Schwarz,
Phys.Lett. {\bf B360} (1995) 13; Erratum-ibid. {\bf B364} (1995) 252,
 {hep-th/9508143} .


 \bibitem{HW} P.~Ho\v rava and E.~Witten,
Nucl. Phys. {\bf
B460} (1996) 506, {hep-th/9510209}.

\bibitem{orientifolds}

G.~Pradisi and A.~Sagnotti,
 Phys. Lett. {\bf B216} (1989) 59;

M.~Bianchi and A.~Sagnotti, 
Phys. Lett. {\bf B247} (1990) 517; {Nucl. Phys. {\bf B361} (1990) 519;

P.~Ho\v rava,  Nucl. Phys. {\bf B327} (1989) 461};
  Phys. Lett. {\bf B231} (1989) 251.


\bibitem{Bergshoeff:1998re} E.~Bergshoeff, E.~Eyras, R.~Halbersma, J.P.~van der
Schaar, C.M.~Hull and Y.~Lozano,
hep-th/9812224.


\bibitem{Hulm} C.M.~Hull,
 JHEP {\bf 9811}  (1998)  27,  {hep-th/9811021} .


\bibitem{strong} C.M.~Hull,
Nucl.\ Phys.\ {\bf B468} (1996) 113 hep-th/9512181.



\bibitem{mal}{J. Maldacena,   hep-th/9711200.}

\bibitem{mats}     T.~Banks, hep-th/9710231;\\
   L.~Bigatti and L.~Susskind, hep-th/9712072

\bibitem{sig} C.M.~Hull,
JHEP {\bf 11} (1998) 017 hep-th/9807127.


\bibitem{fhrs}{W. Fischler, E. Halyo, A. Rajaraman and L. Susskind,
Nucl.Phys. {\bf B501}  (1997) 409,
 hep-th/9703102.}




 \bibitem{bor}{P. Goddard, \lq The Work of R.E. Borcherds', to appear in {\it
Proceedings of the International Congress of
Mathematicians}, and references therein.}



\bibitem{HJ}{C. M. Hull  and B. Julia, Nucl. Phys. {\bf B534} (1998) 250,
hep-th/9803239.}

\bibitem{CPS}{E. Cremmer,  I.V. Lavrinenko,  H. Lu, C.N. Pope,  K.S. Stelle and
 T.A. Tran,
Nucl.Phys. {\bf B534} (1998) 250,
hep-th/9803259.}

\bibitem{Stelle}{ K. S. Stelle, hep-th/9803116.}



\bibitem{Moore} {G. Moore, hep-th/9305139,9308052.}


\bibitem{ebr} C.M.~Hull,
hep-th/9806146.



\bibitem{khur} C.M.~Hull and R.R.~Khuri,
Nucl.\ Phys.\ {\bf B536} (1998) 219 hep-th/9808069.

\bibitem{khura}
C.M.~Hull and R.R.~Khuri,
hep-th/9911nnn.




\bibitem{Hull} C.M.~Hull,
Nucl. Phys. {\bf B509} (1998) 216,
 {hep-th/9705162}.

  \bibitem{to1} P.K.~Townsend,  {hep-th/9712004} .


\bibitem{Julia} B.~Julia, in {\em Superspace and Supergravity},
  eds. S.W.~Hawking and M.~Rocek (Cambridge University Press, 1981)



\bibitem{eten} B.~Julia, in {\em Lectures in Applied Mathematics},
   Vol. {\bf 21} (1985) 355;\\
 H.~Nicolai, {\em Phys. Lett.} {\bf B276} (1992) 333





\bibitem{nic} B.~de Wit, H.~Nicolai,
{\em Phys. Lett.} {\bf 155B} (1985) 47; {\em Nucl. Phys.} {\bf B274} (1986) 363
\bibitem{nic1} H.~Nicolai, {\em Phys.Lett.} {\bf 187B} (1987) 363.
\bibitem{nic2}  H.~Nicolai, hep-th/9801090.


\end{thebibliography}
\end{document}